\documentclass[final,conference,10pt,table,a4paper]{IEEEtran}

\usepackage[utf8]{inputenc}
\usepackage{graphicx}

\usepackage[acronym]{glossaries}
\usepackage{amsmath}
\usepackage{xcolor}
\usepackage{amsfonts}
\usepackage{amssymb}
\usepackage{verbatim}
\usepackage{multicol}
\usepackage{stfloats}
\usepackage{float}
\usepackage{caption}
\usepackage{subcaption}

\usepackage{csquotes}

\usepackage{pdfpages}
\usepackage{geometry}
\usepackage{array}
\usepackage{booktabs}

\usepackage{amsmath}
\usepackage{calc}

\newacronym[plural=RISs, firstplural=reconfigurable intelligent surfaces (RISs)]{ris}{RIS}{reconfigurable intelligent surface}
\newacronym{em}{EM}{electromagnetic}
\newacronym{agv}{AGV}{automated ground vehicle}
\newacronym{iot}{IoT}{internet of things}
\newacronym{6g}{6G}{sixth-generation}
\newacronym{qos}{QoS}{Quality of Service}
\newacronym{uwb}{UWB}{ultra-wideband}
\newacronym{mmwave}{mmWave}{millimetre-wave}
\newacronym{rssi}{RSSI}{Received Signal Strength Indicator}
\newacronym{tdoa}{TDoA}{Time Difference of Arrival}
\newacronym{toa}{ToA}{Time of Arrival}
\newacronym{aoa}{AoA}{Angle of Arrival}
\newacronym{crb}{CRB}{Cramer-Rao lower bound}
\newacronym{3d}{3D}{three-dimensional}
\newacronym{slam}{SLAM}{Simultaneous Localization and Mapping}

\IEEEoverridecommandlockouts

\usepackage[utf8]{inputenc}
\usepackage[english]{babel}

\usepackage[
backend=biber, 
natbib=true, 
style=numeric, 
citestyle=numeric, 
sorting=none
]{biblatex}

\begin{document}

\setlength{\textfloatsep}{3pt}

\title{AI-Assisted NLOS Sensing for RIS-Based Indoor Localization in Smart Factories}

\author{
    \IEEEauthorblockN{
        Taofeek A.O. Yusuf\IEEEauthorrefmark{1}\IEEEauthorrefmark{2},  
        Sigurd S. Petersen\IEEEauthorrefmark{2},
        Puchu Li\IEEEauthorrefmark{2}, 
        Jian Ren\IEEEauthorrefmark{3}, 
        Placido Mursia\IEEEauthorrefmark{1}, \\
        Vincenzo Sciancalepore\IEEEauthorrefmark{1},
        Xavier Costa P\'erez\IEEEauthorrefmark{1}\IEEEauthorrefmark{5},
        Gilberto Berardinelli\IEEEauthorrefmark{2}, and Ming Shen\IEEEauthorrefmark{2}
    }
    \IEEEauthorblockA{
        \IEEEauthorrefmark{1}NEC Laboratories Europe, 69115 Heidelberg, Germany\\ \IEEEauthorrefmark{2}Aalborg University, Fredrik Bajers Vej 7K, 9220 Aalborg Øst, Denmark\\
        \IEEEauthorrefmark{3}Xidian University, Xián, China\\
        \IEEEauthorrefmark{5}i2cat Foundation and ICREA, 08034 Barcelona, Spain\\
    }
}

\maketitle

\IEEEpeerreviewmaketitle







\begin{abstract}
In the era of Industry 4.0, precise indoor localization is vital for automation and efficiency in smart factories. Reconfigurable Intelligent Surfaces (RIS) are emerging as key enablers in 6G networks for joint sensing and communication. However, RIS faces significant challenges in Non-Line-of-Sight (NLOS) and multipath propagation, particularly in localization scenarios, where detecting NLOS conditions is crucial for ensuring not only reliable results and increased connectivity but also smart factory personnel's safety. This study introduces an AI-assisted framework employing a Convolutional Neural Network (CNN) customized for accurate Line-of-Sight (LOS) and NLOS classification to enhance RIS-based localization using measured, synthetic, mixed-measured, and mixed-synthetic experimental data, that is, original, augmented, slightly noisy, and highly noisy data, respectively. Validated through such data from three different environments, the proposed customized-CNN (cCNN) model achieves {95.0\%-99.0\%} accuracy, outperforming standard pre-trained models like Visual Geometry Group 16 (VGG-16) with an accuracy of {85.5\%-88.0\%}. By addressing RIS limitations in NLOS scenarios, this framework offers scalable and high-precision localization solutions for 6G-enabled smart factories.

\end{abstract} 
\vspace{-0.1cm}
\section{Introduction}
The increasing complexity of smart factory environments has necessitated the development of precise indoor localization systems to enable real-time tracking, efficient workflow optimization, and enhanced automation. Traditional methods relying on Line-of-sight (LOS) communication often suffer from accuracy degradation in the presence of obstacles, leading to an increased focus on overcoming Non-Line-of-sight (NLOS) challenges. Recent Artificial Intelligence (AI) advancements have shown immense potential in addressing these issues by leveraging complex data-driven approaches to model the intricate interactions between signals and the environment. 

While the global vision of this study is the Localization in Smart Factories with mobile Reconfigurable Intelligent Surfaces (mobile-RIS), this current study focuses on integrating Artificial Intelligence (AI) with the sensing capability of RIS to address LOS and NLOS signal propagation challenges within Smart factories. The proposed customized-CNN (cCNN) model architectural framework aims to improve sensing accuracy in dynamic and cluttered environments through training on real-world measured, synthetically generated, and noisy-mixed-synthetic datasets in different environments. The presented cCNN architecture for LOS/NLOS classification is composed of a feature extraction and classification module with its design prioritizing accurate feature learning and efficient classification tailored to sensing and propagation scenarios. The feature extraction involves sequential 2D convolutional layers with Rectified Linear Unit (ReLU) activation, followed by Batch Normalization and pooling operations to extract spatial patterns from the input Channel Impulse Response (CIR)/Spectrogram image data. The first two layers progressively produce deeper feature maps, followed by a flattening layer that transforms the extracted features into fully connected layers within the classification module. Additionally, a SoftMax activation function is utilized to generate the final classification probabilities for LOS, while NLOS is introduced using metal plates measuring 1m × 1m and 0.75m × 0.75m, respectively. 



\vspace{-0.1cm}
\section{Related Works}
In the realm of indoor localization, integrating LOS and NLOS sensing and communication remains a critical research area. Recent advancements, such as low-complexity estimators for RIS-assisted localization, address issues like unknown RIS positions and orientations, achieving high accuracy even in complex Near-Field (NF) scenarios  \cite{5}. Further innovations include atomic norm minimization for single-snapshot localization, enabling efficient parameter estimation, and beam squint mitigation algorithms, both of which enhance accuracy under synthetic and real-time data conditions \cite{6,7}. The concept of Integrated Sensing and Communication (ISAC) and RIS in 6G systems has gained traction, offering simultaneous localization and communication while requiring optimized algorithms for resource allocation \cite{8,17}. Multi-user RIS frameworks and Extra-Large RIS (XL-RIS) systems have expanded the scope of localization, tackling challenges like hybrid-field beam squint and dynamic channel sensing in high-mobility environments \cite{9,10,11,12,18,19}.

Efforts to unify RIS and ISAC have led to innovative frameworks, such as nested tensor-based models and Continuous Intelligent Surfaces (CIS), improving scalability and adaptability in dynamic conditions \cite{13,14}. Joint optimization of RIS configurations has proven effective in balancing communication and positioning performance, as shown in low-complexity schemes for RIS-aided positioning \cite{15,20}. Practical advancements in phase shift models further enhance real-world localization accuracy by optimizing directional and positional phase profiles \cite{16}. Meanwhile, side-link communication approaches have successfully converted NLOS to LOS conditions, leveraging IoT devices as relay nodes to enhance area coverage and reduce outage probabilities in factory environments \cite{2}.

RIS-assisted Integrated Sensing and Backscatter Communication (ISABC) systems address power limitations in IoT devices, improving energy efficiency and spectral utilization \cite{1}. LiDAR-assisted RIS systems have also emerged as a transformative approach, combining LiDAR's precise spatial mapping with RIS's adaptability to overcome multipath-induced errors. These hybrid models significantly enhance localization accuracy and open pathways for scalable implementations in IoT networks, smart cities, and autonomous systems \cite{3,4}. Together, these advancements underscore the immense potential of RIS technologies in transforming smart factory sensing, localization and communication operations. Through these interconnected advancements, the vision of AI-assisted LOS and NLOS sensing for indoor localization in smart factories is steadily becoming a reality, with RIS technology playing a central role in enhancing both localization precision and communication performance in complex industrial environments.

Most of the studies done thus far were based on synthetically generated data and, in some cases, real-world data with little actual experimentation. In our study, we have exploited experimental data obtained from three different real-world scenarios, namely i) a controlled environment (i.e., the Anechoic Chamber), ii) a moderately-noisy environment (i.e., the Meeting Room) and iii) a highly-noisy environment (i.e., the High-Frequency (HF) Lab) and AI to assist RIS in sensing its environment in search of possible obstacles, i.e., NLOS conditions. 


\section{Measurement Campaign and Architecture}
The measurement campaign aims to gather experimental data from three distinct scenarios with three different complexities to support AI training in differentiating between LOS and NLOS scenarios. Each scenario involves an RIS, two metal plates of various sizes to create an NLOS blockage scenario for the experiment, and a turntable. The RIS utilizes its beam-steering capability to redirect the signal path toward the receiver in the NLOS experimental scenario. Additionally, the turntable allows scanning the environment to identify and analyze multipath propagation. The equipment used for conducting the experimental measurements is given in Table I.

\begin{table}[h]
    \centering
    \caption{Equipments used for conducting the experimental measurements.}
    \label{tab:colored_table}
    \resizebox{0.50\textwidth}{!}{
        \begin{tabular}{|c|c|c|}
        \hline
        \textbf{Devices and parts} & \textbf{Name} & \textbf{Specification} \\ \hline
        VNA     & Keysight PNA N5227B     &  900 Hz to 67 GHz  \\ \hline
        Metal plate 1    & 75 cm x 75 cm & 5 mm thick     \\ \hline
        Metal plate 2     & 100 cm x 100 cm & 5 mm thick     \\ \hline
        Mechanical turntable    & HRT-I-Turntable & 360 deg, 0.1 deg step \\ \hline
        Rx horn Antenna    &  Own design   & 4.3 GHz - 7 GHz     \\ \hline
        Tx horn Antenna    & Own design    & 4.3 GHz - 7 GHz     \\ \hline
        RF amplifier    & ZVE-8G    & 2 - 8 GHz, 30 dB Gain     \\ \hline
        \end{tabular}
    }
\end{table}


\begin{figure}[h]
    \centering
    \includegraphics[width=1.0\linewidth, angle=0, trim=0 0 0 0, clip]{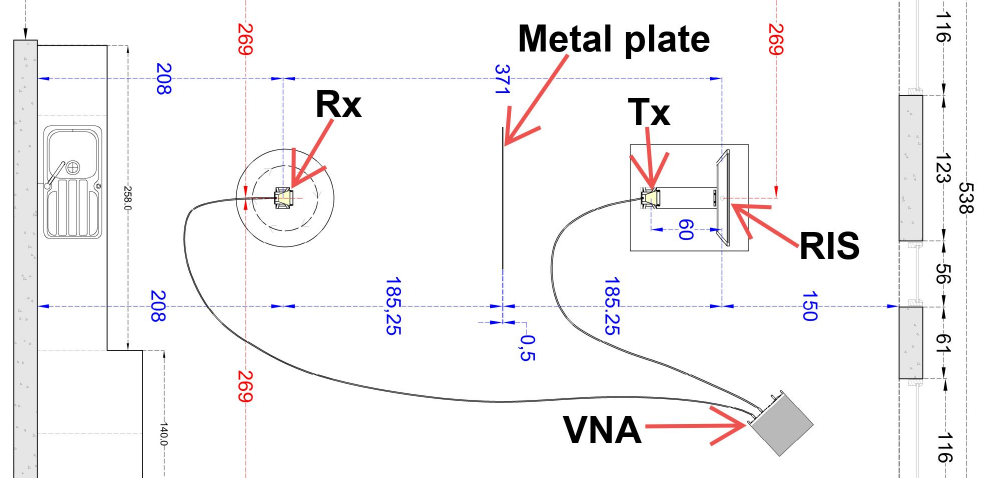}
    \caption{Meeting Room's LOS and NLOS experimental measurements with azimuth, RIS's elevation, and the Rx-horn-antenna's direction all set to zero.}
    \label{fig:Kitchen}
\end{figure}

\begin{figure}[h]
    \centering
    \includegraphics[width=0.9\linewidth, trim=50 0 60 0, clip]{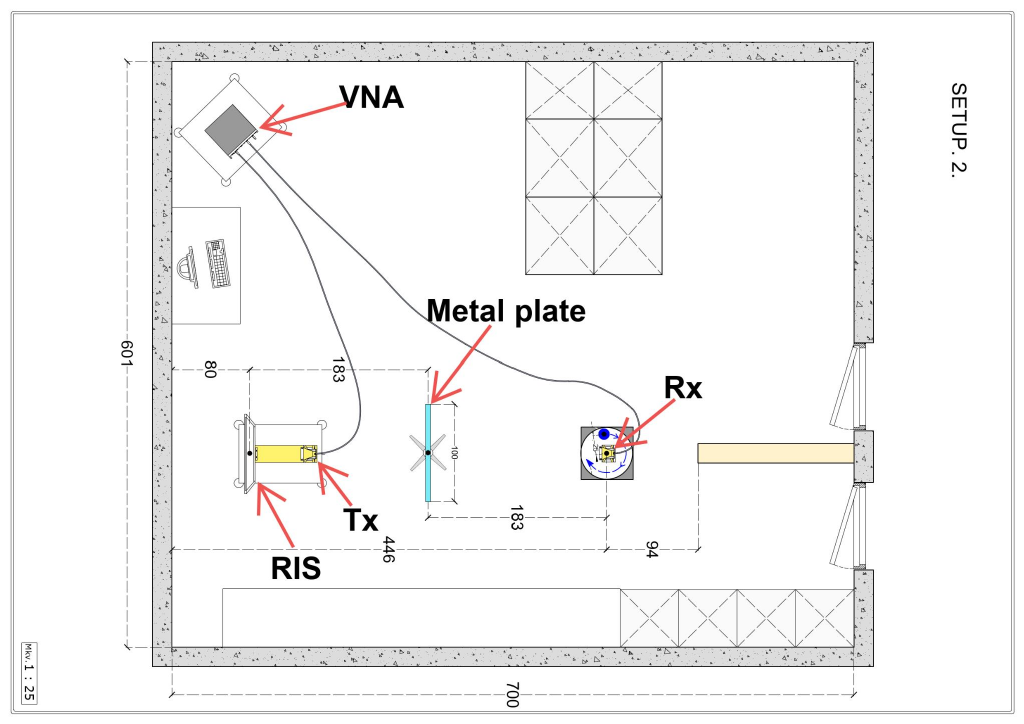}
    \caption{HF lab's LOS and NLOS  experimental measurements with azimuth, RIS's elevation, and the Rx-horn-antenna's direction all set to zero.}
    \label{fig:Chamber}
\end{figure}

\begin{figure}[h]
    \centering
    \includegraphics[width=0.85\linewidth, trim=150 150 150 150, clip]{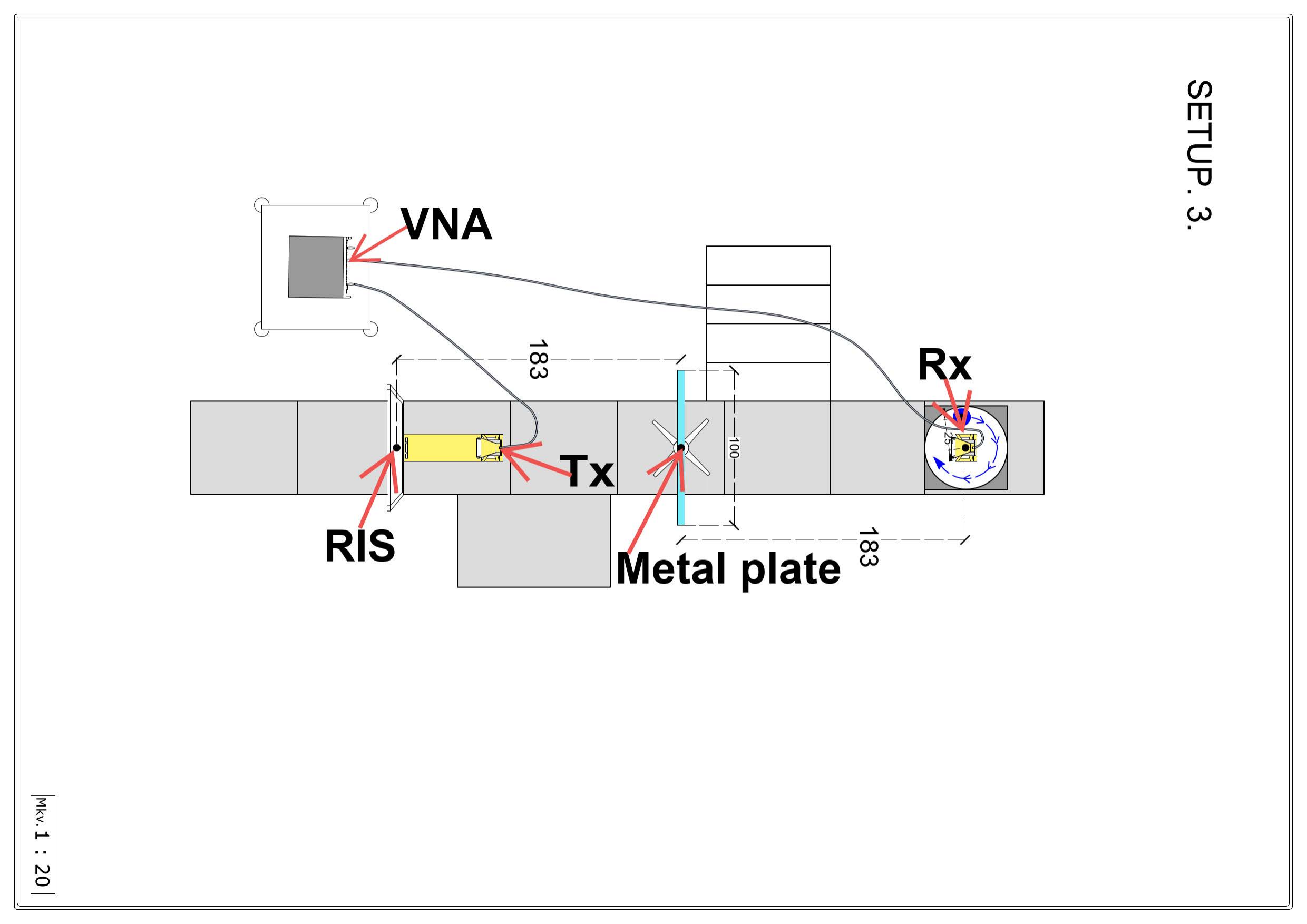}
    \caption{Anaechoic chamber's LOS and NLOS  experimental measurements with azimuth, RIS's elevation, and the Rx-horn-antenna's direction all set to zero.}
    \label{fig:Chamber}
\end{figure}

\begin{figure*}[t]
    \centering
    \includegraphics[width=1.0\linewidth, angle =0, trim=10 190 5 10, clip, clip]{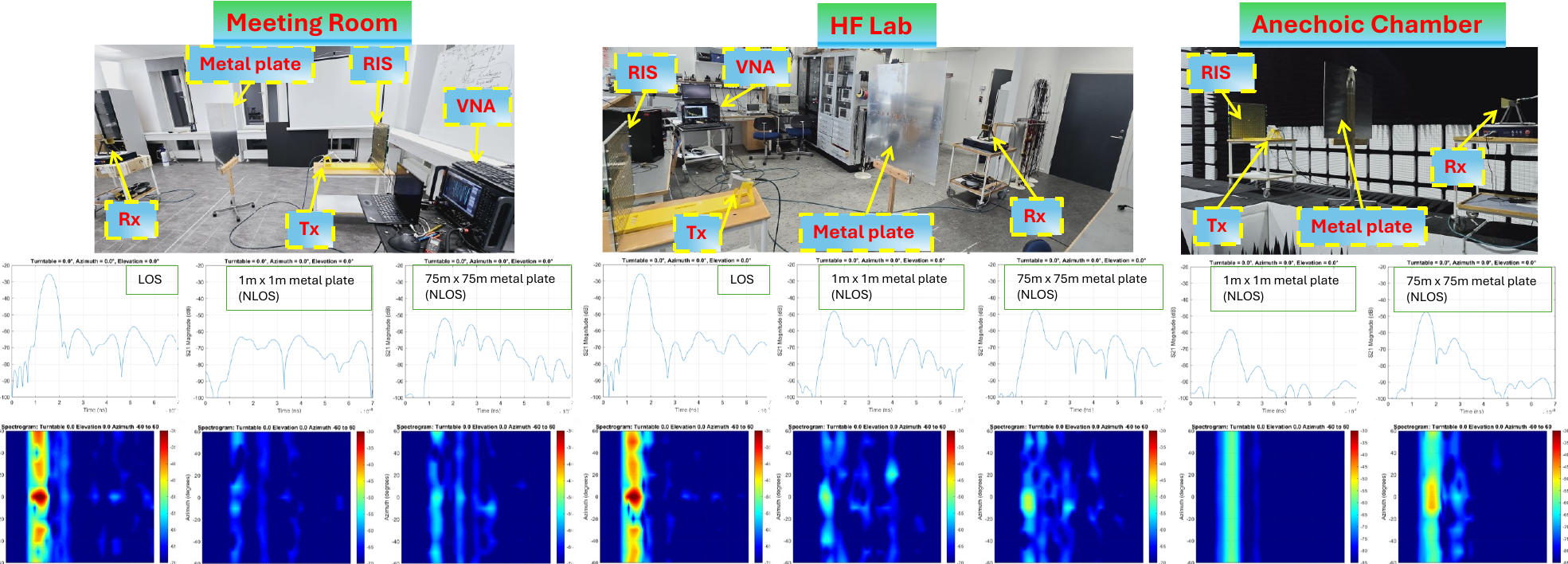}
    \caption{CIR and spectrogram image data experimental measurement scenarios for LOS and NLOS using 5mm thick 1m-by-1m and 0.75m-by-0.75m metal-plates blockade to create the NLOS scenarios, respectively.}
     \label{fig:Measurement_R}
\end{figure*}

\subsubsection{General setup}
The measurement system comprises a Vector Network Analyzer (VNA) operating between 4.8 GHz and 5.2 GHz, a 16-by-16 RIS panel, a transmitter, and a receiver mounted on a turntable, as seen in Figures [1-4] where the VNA port-1 is connected to the transmitter in front of the RIS panel. In contrast, port-2 is connected to the receiver. The turntable rotates 360$^{\circ}$ with a 5$^{\circ}$ increment, allowing for spatial characterization of the environment. The RIS panel is fixed at 0$^{\circ}$ of azimuth and elevation for all scenarios. Channel Impulse Response (CIR) data were recorded at each rotational 5$^{\circ}$ step and stored to serve as data for the AI model training. The measurements were conducted in the Meeting Room, High-Frequency (HF) Lab, and Anechoic Chamber at the Aalborg University Antenna Laboratory, respectively.

\subsubsection{Scenario 1: Experimental Measurement for Meeting Room Environment Scenario}
Figure \ref{fig:Kitchen} depicts the sketch of the meeting room environment where the first set of measurements was taken. The essence of this measurement is to portray part of the smart factory scenario where the environment is moderately crowded with factory equipment, with few multipath signal rebounds and some high shelves that could be blocking the signal path.  The distance between the receiver and the transmitter is 0.431m, with a metal plate positioned between them to construct the NLOS scenario. The Meeting room’s reflective surfaces introduced multi-path effects, providing a realistic indoor dataset for AI model development.

\subsubsection{Scenario 2: Experimental Measurement for HF Laboratory Environment Scenario}
The second scenario was conducted in an HF laboratory to emulate a smart factory environment, where numerous reflective surfaces caused increased environmental complexity and dense multi-path effects with the presence of the respective blockades to emulate the NLOS condition. The transmitter emits signals toward the RIS and redirects them to the receiver on the turntable. The configuration remained consistent with the meeting room setup, allowing comparison across scenarios.

\subsubsection{Scenario 3: Experimental Measurement for Anechoic Chamber Environment Scenario}
The third scenario utilized an anechoic chamber to minimize reflections and eliminate multipath effects, providing a baseline for characterizing LOS and NLOS conditions under close to ideal circumstances. The RIS redirected the transmitted signals, and CIR data were recorded at the receiver antenna at each 5$^{\circ}$ increment position of the turntable.

\begin{figure*}[t!]
    \centering
    \includegraphics[width=5.8in, angle=0, trim=15 145 2.5 25, clip]{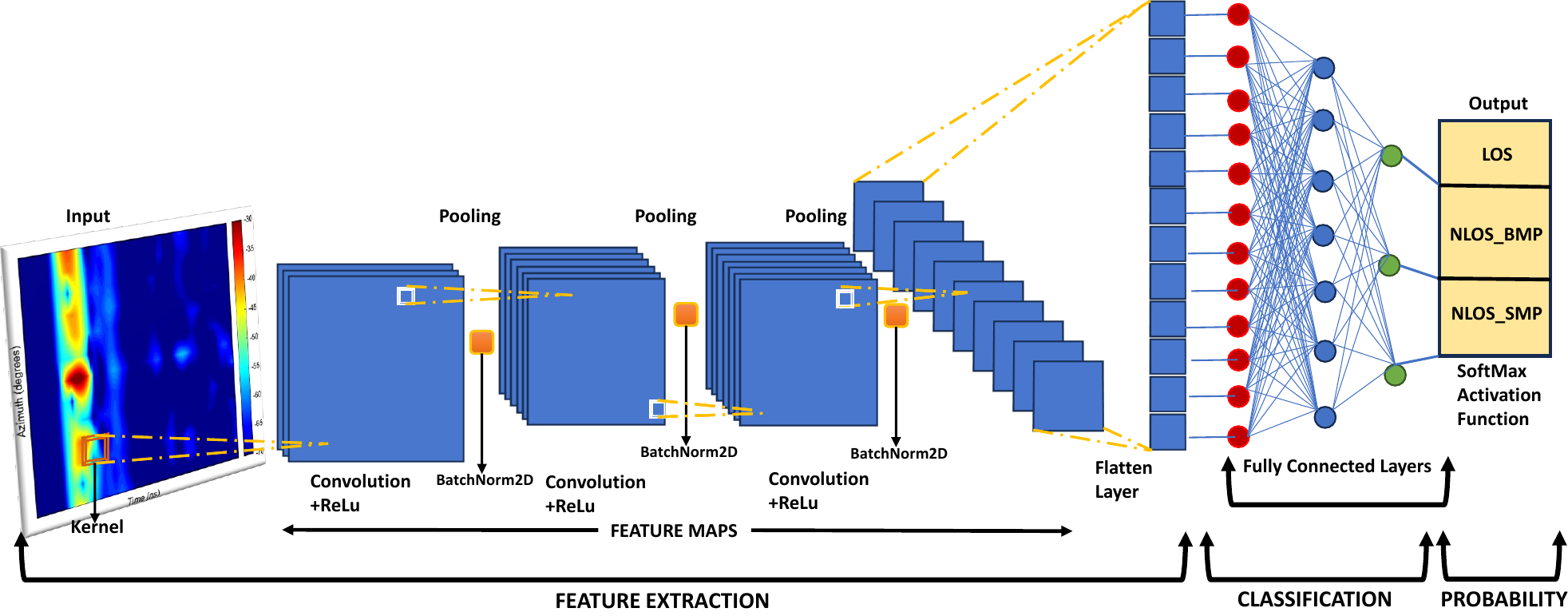}
    \caption{Deep learning architecture for AI-assisted LOS and NLOS sensing.}
    \label{fig:AI-architecture}
\end{figure*}

The measurement in Figure \ref{fig:AI-architecture} illustrates the results under various conditions, including LOS, NLOS with 1m-by-1m blockage, and NLOS with 0.75m-by-0.75m blockage. The experimental data are represented in CIR and spectrograms, which serve as input for AI model training. These measurements are crucial as they originate from diverse real-world scenarios, ensuring the robustness and applicability of the AI system to varied environmental conditions.

Using distinct environments and blockage sizes underscores the importance of capturing a wide range of propagation characteristics. This diversity in the training dataset enhances the AI model's ability to generalize and perform effectively in our current study and experimental application of AI-assisted RIS-based indoor sensing and possible localization in dynamic smart factory settings. Based on our real-world measurement scenarios, this study ensures that the AI's learning process reflects realistic operational challenges, enhancing its performance in realistic smart factory deployment.  


\section{AI-Assisted Sensing for LOS and NLOS Classification}

The customized CNN designed for the AI-assisted NLOS sensing scenarios sequentially processes the measured CIR-analysed Spectrogram PNG input image data captured during the scenario experiments in Section II, transforming it through various layers to achieve the final AI-Assisted Sensing classification for the experimental measured data. The entire CNN model can be expressed as a composite function, such that, 
\vspace{-0.2cm}
\begin{multline}\label{eq:1}
    f(X) = \text{softmax}\Big(W_{\text{fc2}} \cdot \text{dropout}\Big(
            \sigma\Big(W_{\text{fc1}} \cdot \text{flatten}\big(g(X)\big) \\
            + b_{\text{fc1}}\Big)\Big) + b_{\text{fc2}}\Big)
\end{multline}

where g(X) represents the sequential application of convolutional, batch normalization, activation, and pooling layers, and it is given as:
\vspace{-0.2cm}
\begin{equation}\label{eq:2}
\begin{split}
    g(X) &= \text{Pool3}\Big( \sigma\Big( \text{BN3}\Big( W_3 \ast \text{Pool2} \Big( \sigma\Big(\text{BN2} \Big( W_2 \ast \text{Pool1}\\
         &\quad   \Big( \sigma\Big( \text{BN1} \Big( W_1 \ast X + b_1 \Big) \Big) \Big) \Big) \Big) \Big) \Big) \Big). 
\end{split}
\end{equation}

The input to the CNN is a three-channel image of size \( 224 \times 224 \), and it can be represented as $X_{\text{in}} \in \mathbb{R}^{3 \times 224 \times 224}$. Each pixel is normalized, and the input is represented as a tensor, where $W_1 \in \mathbb{R}^{32 \times 3 \times 3 \times 3}$ are the convolutional filters, $b_1 \in \mathbb{R}^{32}$ are the biases, $\ast$ denotes the convolution operation, $\sigma$ is the ReLU activation function. A max-pooling operation follows this to reduce the spatial dimensions by half, such that: 
\vspace{-0.1cm}
\begin{equation} \label{eq:3}
    \resizebox{0.50\textwidth}{!}{$ 
        X_{\text{pool1}} = \text{MaxPool}(X_{\text{conv1}}, \text{kernel\_size}=2, \text{stride}=2)
    $}
\end{equation} 

and the output dimensions after this block are $32 \times 112 \times 112$. The second convolutional block applies 64 filters of size $3 \times 3$, followed by batch normalization, ReLU activation, and max-pooling thus:
\begin{flalign} \label{eq:4}
    X_{\text{conv2}} &= \sigma \left( \text{BatchNorm} \left( W_2 \ast X_{\text{pool1}} + b_2 \right) \right) & \\
    X_{\text{pool2}} &= \text{MaxPool}(X_{\text{conv2}}, \text{kernel\_size}=2, \text{stride}=2). &
\end{flalign}
\vspace{-0.1cm}
and the output dimensions after this block are $( 64 \times 56 \times 56)$. The third convolution block contains $128$ filters of size $3 \times 3$ that are applied, followed by the same sequence of operations as in the earlier convolution block such that: 
\vspace{-0.2cm}
\begin{flalign}\label{eq:5}
    X_{\text{conv3}} &= \sigma \left( \text{BatchNorm} \left( W_3 \ast X_{\text{pool2}} + b_3 \right) \right) \\
    X_{\text{pool3}} &= \text{MaxPool}(X_{\text{conv3}}, \text{kernel\_size}=2, \text{stride}=2)
\end{flalign}
where the output dimensions after this block are $128 \times 28 \times 28$.

The output of the final convolutional block is flattened into a one-dimensional vector to prepare it for the fully connected layers thus:
\vspace{-0.1cm}
\begin{equation} \label{eq:6}
    X_{\text{flatten}} = \text{Flatten}(X_{\text{pool3}})
\end{equation} 

 which transforms the dimensions from $128 \times 28 \times 28$ to $100,352$ and the first fully connected layer reduces the flattened vector to $256$ dimensions, where: 
 \vspace{-0.2cm}
 
 \begin{equation} \label{eq:7}
    X_{\text{fc1}} = \sigma \left( W_{\text{fc1}} X_{\text{flatten}} + b_{\text{fc1}} \right) 
\end{equation} 

 $W_{\text{fc1}} \in \mathbb{R}^{256 \times 100,352}$ and $b_{\text{fc1}} \in \mathbb{R}^{256}$. A dropout layer with a probability of $p=0.5$ is applied to prevent overfitting, 
\vspace{-0.1cm}
 \begin{equation} \label{eq:8}
    X_{\text{dropout}} = \text{Dropout}(X_{\text{fc1}}, p=0.5)
\end{equation}  

 and the final fully connected layer maps the $256$-dimensional vector to the number of classes, $3$, having:
\vspace{-0.4cm}
  \begin{equation} \label{eq:9}
    X_{\text{output}} = \text{Softmax}(W_{\text{fc2}} X_{\text{dropout}} + b_{\text{fc2}})  
\end{equation} 
where $W_{\text{fc2}} \in \mathbb{R}^{3 \times 256}$ and $b_{\text{fc2}} \in \mathbb{R}^{3}$. The final output is a probability distribution over the three classes, $X_{\text{output}} \in \mathbb{R}^3, \quad \text{such that} \quad \sum_{i=1}^3 X_{\text{output}, i} = 1$. \\

\begin{table*}[ht]
    \centering
    \caption{Comparison of ML Methods in Multiple Scenarios with Different Data Types and Sizes.}
    \label{tab:benchmark_accuracy}
    \begin{tabular}{@{}lccccccc@{}}
    \toprule
    \textbf{ML Methods} & \multicolumn{3}{c}{\textbf{CNN}} & \multicolumn{3}{c}{\textbf{VGG-16}} \\ 
    \cmidrule(lr){2-4} \cmidrule(lr){5-7}
    \textbf{Training Scenarios} & \textbf{Meeting Room} & \textbf{HF Lab} & \textbf{Chamber} & \textbf{Meeting Room} & \textbf{HF Lab} & \textbf{Chamber} \\ 
    \midrule
    \midrule
    \multicolumn{7}{l}{\textbf{Data Types|Set Size}} \\ 
    Measured Data|(72)        & \textbf{95.0\%} & 86.0\% & \textbf{99.9\%} & 71.0\% & 29.0\% & 64.0\% \\ 
    Synthetic Data|(720)    & 93.0\% & 92.0\% & 94.0\% & 50.0\% & 38.0\% & 51.2\% \\ 
    Mixed Measured Data|(144)      & 94.0\% & \textbf{93.0\%} & 94.0\% & 86.0\% & 75.0\% & 86.0\% \\ 
    Mixed Synthetic Data|(725)       & 88.0\% & 86.0\% & 88.0\% & \textbf{88.0\%} & \textbf{85.5\%} & \textbf{88.0\%} \\ 
    \bottomrule
    \end{tabular}
\end{table*}




\begin{figure*}[t]
    \centering
        \subfloat[\centering Customized CNN Accuracy Scores]{{\includegraphics[width=0.47\textwidth]{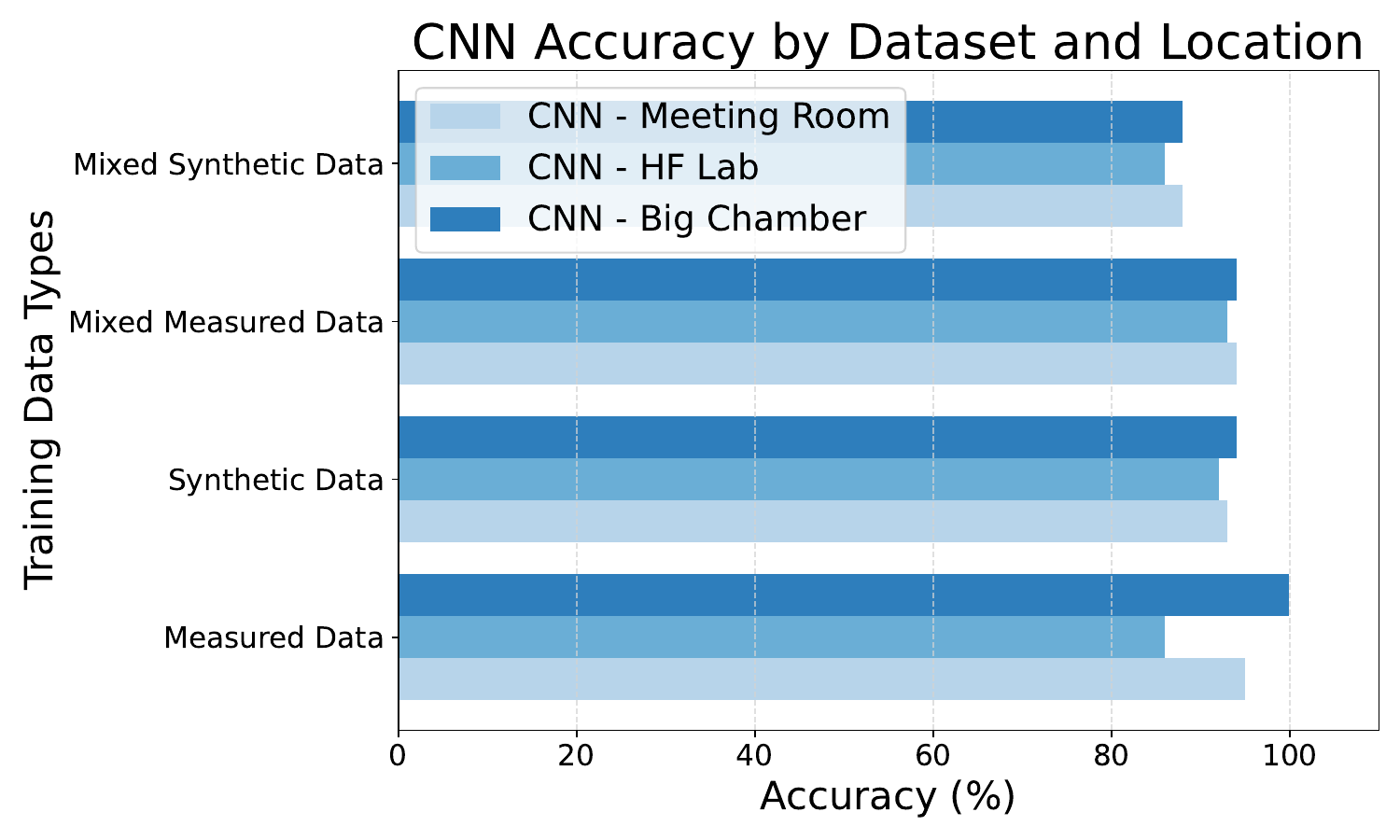}}}
        \qquad
        \subfloat[\centering Pre-Trained VGG-166  Accuracy Scores]{{\includegraphics[width=0.47\textwidth]{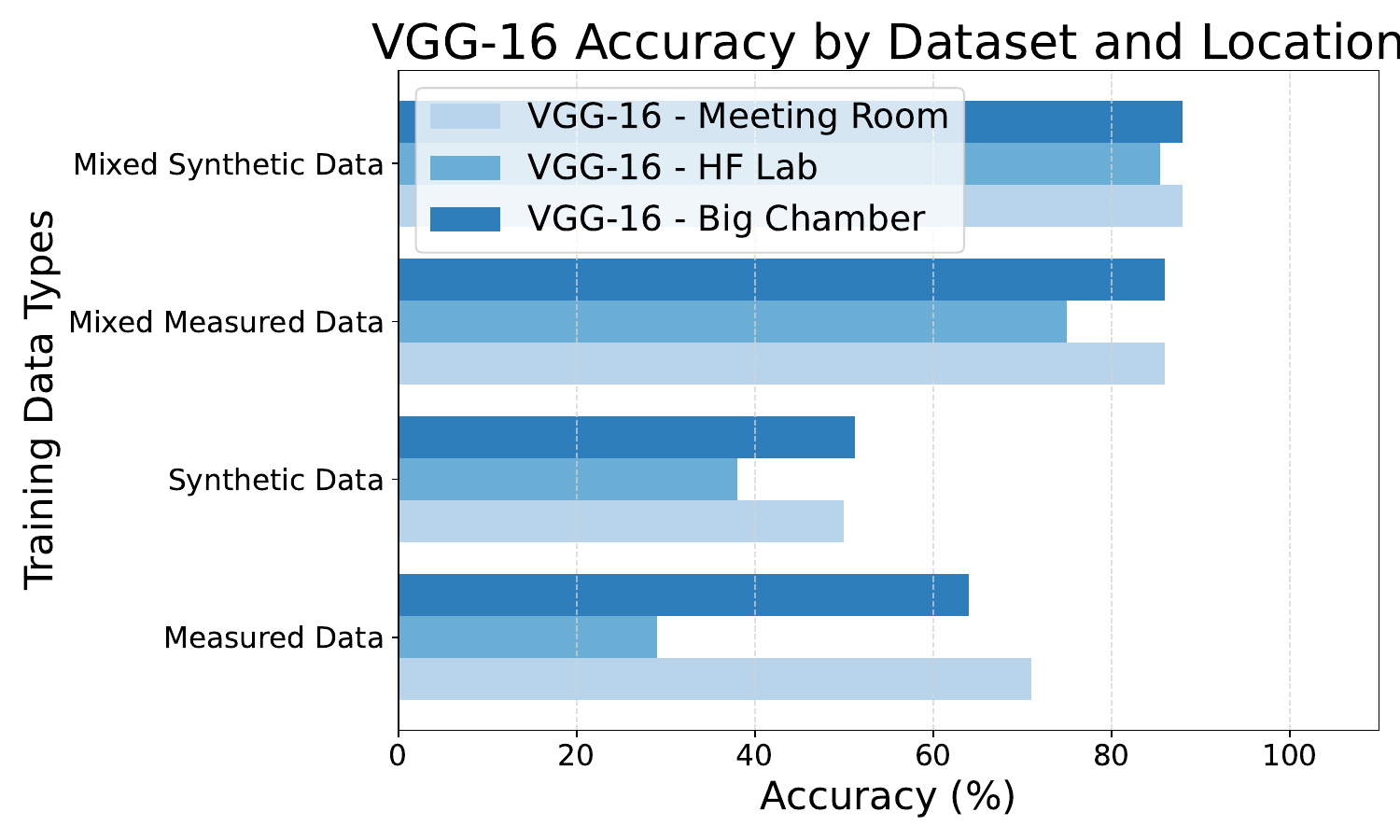}}}
        \caption{Model's accuracy scores for different environmental scenarios and data genres.}
        \label{fig:example}
\end{figure*}

\vspace{-0.5cm}
\section{ Inferences and Analysis}
The CNN architecture in Figure 5 processes PNG Spectrogram input image datasets through three convolutional blocks, each comprising convolution, batch normalization, ReLU activation, and max-pooling layers, which progressively extract hierarchical features while reducing spatial dimensions. The extracted features are flattened and passed through fully connected layers with dropout regularization to prevent overfitting, and a softmax classifier outputs the probability distribution over the three target classes. The CNN model was trained on four variations of the dataset namely, measured data, synthetic data, mixed measured data, and mixed synthetic data, across three distinct environments, listed as a Meeting Room, an HF Laboratory, and an Anechoic Chamber with resulting outputs across distinct environments presented in Table II and Figure 6, along with their analysis, respectively.

\subsection{Experimental Datasets (Original and Generated)}
\subsubsection{Measured (a.k.a. Original) Dataset}
These are the original experimental measured data as specified in the defined scenarios in Section III. These CIR/spectrogram image data were carefully recorded using the experimental setup of different areas' conditions of the smart factory environment.

\subsubsection{Synthetic (a.k.a. Augmented) Dataset}
The Synthetic data was generated using a sample of each of the LOS and NLOS depicted with a metal-plate of size 1m-by-1m and 0.75m-by-0.75m block, respectively, of the original dataset and synthesized using some transformative operations such as flipping, rotation, resizing, and cropping in addition to tweaking the saturation, brightness, contrast and hue of the original datasets.

\subsubsection{Mixed Measured (a.k.a. Slightly-noisy) Dataset}
This dataset is a mixture of the original dataset with the slightly noisy synthetic dataset in a ratio that permits more of the original than the slightly-noisy synthetic one to test the model's strength when datasets that are not up to par are used for the AI model training.

\subsubsection{Mixed Synthetic (a.k.a. Highly-noisy) Dataset}
To further test the adaptability and performance of the AI model, we made the synthetic data noisier and mixed it in a ratio that permits more of the noisy-synthetic data than the original measured dataset. This is to provide usage in a more reliable
and well-defined propagation environment and make it suitable for data-scarce applications for AI model training.

\subsection{Meeting Room}
The cCNN model demonstrated a strong ability to interpret real-world signals, achieving 95.0\% accuracy on Experimental Measured (original) data in a controlled environment as compared to the VGG-16 model with 71.0\% accuracy, and its performance accuracy on synthetic data is 93.0\% compared to the VGG-16 model at 50.0\% accuracy. The mixed-measured (slightly-noisy) data restored accuracy to 94.0\%, underscoring the benefit of combining real-world and synthetic datasets for robustness as compared to its counterpart, which has a performance accuracy of 86.0\%. With the prior knowledge that the VGG-16 model is a state-of-the-art model that has been tested and trusted over time, we try to compare our customized model to see how well it fares in a scenario where there is an original data-scarce situation. Our model and the VGG-16 model achieved an accuracy of 88.0\% on the Mixed Synthetic (highly-noisy) data, depicting how robust our model is to learn on non-original data, indicating its utility when original measured data is limited or not available. 

\subsection{HF Laboratory}
In the HF lab environment, our cCNN model achieved an 86.0\% accuracy on experimentally measured (original) data, slightly lower than in the meeting room due to increased complexity and reflections as compared to the VGG-16 model with 29.0\% accuracy. Its performance accuracy on synthetic data increased to 92.0\%, reflecting the power of the model's learning capability. Our model had a performance accuracy of 93.0\% on the Mixed-Measured (slightly noisy) data as compared to the VGG-16 model with a 75\% accuracy, while our customized model, at 86\% accuracy, slightly outperformed the VGG-16 model, at 85.5\% accuracy, on the Mixed-Synthetic (highly noisy) dataset. Again, this showcases that our model is robust and competes with the VGG-16 model in a situation where there is little to no availability of the original measured data with encouraging adaptable performance. 

\subsection{Anechoic Chamber}
In the Anechoic Chamber, which is the largest environment of the experiments, CNN achieved a performance accuracy of 99.9\% accuracy on experimentally measured (original) data as compared to the VGG-16 model with an accuracy of 64.0\%, leveraging the clear distinctions in LOS and NLOS propagation. Our customized model had a 94.0\% performance accuracy on synthetic data, benefiting from the chamber's well-defined multipath characteristics as compared to the VGG-16 model's performance with an accuracy of 51.2\%. The Mixed-measured (slightly noisy) datasets showed stable performance, achieving 94.0\% performance accuracy as compared to the VGG-16 with an accuracy of 86\%. The customized model's performance and that of the VGG-16 model on the Mixed-synthetic (highly noisy) dataset is 88.0\% accuracy, respectively. 

Thus, the customized CNN model outperformed the pre-trained VGG-16 in most scenarios, particularly in the HF Lab and Meeting Room, where real-world signal variability posed challenges. While VGG-16 struggled with measured and synthetic datasets, it performed comparably on mixed synthetic data, making it suitable for situations where there's little to no original measured data or can be used on data-scarce applications.

\section{Conclusion}
This study proposes a robust AI-assisted framework integrating customized CNN architectures with RIS to address the critical challenges of NLOS sensing and detection in smart factory indoor environments towards localization operation. By achieving the AI model's performance accuracy of 95.0\%-99.0\% across diverse real-world scenarios, the proposed solution demonstrates exceptional adaptability, scalability, and robustness, significantly outperforming the existing standard pre-trained model (VGG-16) in complex NLOS and multipath environments. Notably, this work stands out as one of the few studies leveraging extensive experimental data collected in three distinct environments, ensuring the practical applicability and reliability of the proposed approach. Thus highlighting the transformative potential of integrating RIS technology with deep learning models to enhance the precision and reliability of environmental sensing and detection, paving the way for advanced automation and operational efficiency in Industry 4.0. For future work, we aim to refine RIS-based systems by investigating distributed-RIS for LOS and NLOS sensing, detection, and possible localization scenarios. 

\section*{Acknowledgment}
This project has received funding from the European Union’s Horizon 2020 research and innovation program under the Marie Skłodowska-Curie grant agreement No 956670 (5GSmartFact), as well as through Smart Networks and Services Joint Undertaking grant No. SNS- JU-101139130 (6G-DISAC). The authors would also like to thank Omar Cornette (OC-TEKN) for contributing to this paper.





\printbibliography
\end{document}